\documentclass{article}

\PassOptionsToPackage{numbers,sort&compress}{natbib} 

\usepackage[preprint]{style} 

\usepackage[utf8]{inputenc} 
\usepackage[T1]{fontenc}    
\usepackage{hyperref}       
\usepackage{url}            
\usepackage{booktabs}       
\usepackage{amsfonts}       
\usepackage{nicefrac}       
\usepackage{microtype}      
\usepackage{amsmath,amsthm,amsfonts,amssymb,amscd}
\usepackage{mathtools}
\usepackage{lastpage}
\usepackage{enumerate}
\usepackage{fancyhdr}
\usepackage{mathrsfs}
\usepackage[x11names]{xcolor}
\usepackage{graphicx}
\usepackage{listings}
\usepackage{subcaption}
\usepackage[english]{babel}
\usepackage{apxproof}
\usepackage{thmtools}
\usepackage{thm-restate}
\usepackage{enumitem}
\usepackage{float}
\usepackage{bm}
\usepackage{wrapfig}
\usepackage{algorithm}
\usepackage{algpseudocode}

\usepackage{sidecap}
\usepackage{siunitx}

\usepackage[title]{appendix}
\usepackage{outlines}

\theoremstyle{definition}

\definecolor{myGreen}{rgb}{0,0.5,0}

\title{The Standard Model of the Retina}

\author{
  Markus Meister \\
  Division of Biology and Biological Engineering\\
  California Institute of Technology\\
  \texttt{meister@caltech.edu}\\
}

\begin{document}

\maketitle

\section{Abstract}
The scientific study of the retina has reached a remarkable state of completion. We can now explain many aspects of early visual processing based on a relatively simple model of neural circuitry in the retina. The same model, with different parameters, produces a great diversity of neural computations. In this article I lay out what that ``standard model'' is and how it accounts for such a diversity of phenomena. The emergence of such a powerful standard model is unique in systems neuroscience, and I consider what conditions made it possible. The standard model now serves as a baseline from which to organize future retinal research, either by testing the model's assumptions directly, or by identifying phenomena that remain unexplained.

\section{Introduction}
Recently, I had an unusually interesting experience with peer review. 
Three reviewers commented on a new report about visual processing in the retina, and independently all three wrote along the lines of: ``This is exciting, a new phenomenon of early vision, and definitely worth publishing. On the other hand, everything about it can be explained based on the model we have for how the retina works.'' Then the three reviewers laid out how that explanation would run, and how it would predict not only the qualitative observations but also some quantitative details in this new study. 

This episode shows that there exists at least in the minds of many scientists, a standard model of how the retina works. And that this standard model is sufficiently powerful to not only fit the past observations, but also to explain phenomena that haven’t even been reported yet. In this article I want to summarize what that standard model is, and how it manages to cover so much of retinal phenomenology. 
I will explore what were the favorable conditions that allowed the community to come to this state of understanding, compared to other parts of the brain. 
It is also worth considering which concepts or approaches were \emph{not} needed to achieve this level of progress. 
Indeed, this article may be of particular interest to readers whose research focus lies outside the retina.

A few caveats are in order: First, I will treat the retina specifically from the point of view of visual processing. The research challenge is to understand what the retina does with visual signals from photoreceptors based on the structure and function of its neural circuit. There are, of course, other reasons to work on the retina, for example, to understand development of the vascular system or problems of neural degeneration. Those are not the subjects of this article.

Second, this is not intended as an exhaustive review of primary research but a high-level assessment of where our understanding stands in this field and compared to others. I will illustrate the arguments with examples, and liberally cite reviews of the field by others. A great resource for general function and structure of the retina is the online book WebVision~\cite{kolb_webvision_nodate}; regarding visual computations in the retina, I recommend the book ``Retinal Computation''~\cite{schwartz_retinal_2021}.

Finally, I will not claim that everything is understood about the retina, even in this restricted domain. Open questions remain, and I will suggest a few at the end. Nonetheless these research questions are much more sophisticated than those under discussion elsewhere in brain science. 
As someone once said: ``We used to be confused; now we are still confused, but at a much higher level.'' 

\section{Premises}
In college, I learned quantum mechanics from a book that took a fully axiomatic approach~\cite{Cohen-Tannoudji:101367}: 
The text laid out the five postulates of quantum mechanics, and then proceeded to derive all the observable phenomenology from those axioms.
While this may be the ideal final state of a mature science, we have not quite reached it in neurobiology. 
Nevertheless, here are a few postulates that underlie research on the retina and form the basis of the standard model:

1. The purpose of the retina is to make light visible. 
It acts as the interface that converts the physical stimulus -- photons from different directions in space -- into the universal neural language, namely action potentials.
So the goal of the standard model should be to explain and predict the output of the retina -- the firing of retinal ganglion cells -- as a function of the input -- the light patterns on the photoreceptors. 
If we can accomplish that, then we have captured the contribution that the retina makes to human vision.

2. There are about 40 types of retinal ganglion cells. 
Each of these types is the output of a circuit that leads all way from the photoreceptors to the retinal ganglion cell. 
Different types of retinal ganglion cell are distinct by the details of that circuit. 
Each circuit will correspond to a set of parameters of the standard model.

3. In different vertebrate species, the retina follows the same principles, but with different circuit parameters. Different species have undergone different types of selection for these parameters as a result of evolutionary pressure and ecological niches~\cite{hahn_evolution_2023}.

\begin{figure}[H]
    \centering
    \includegraphics[width=0.75\textwidth]{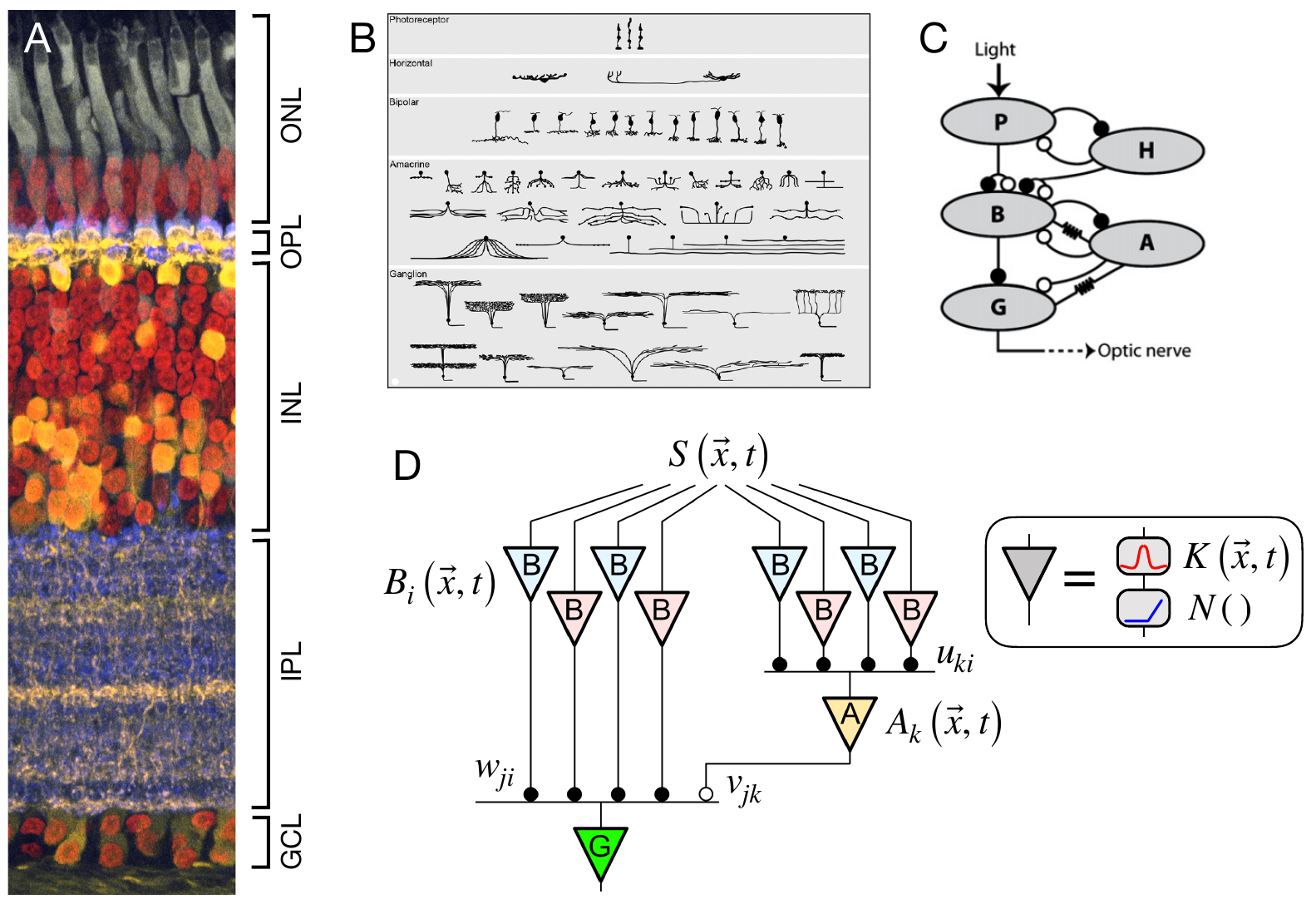}
    \caption{\textbf{Retinal circuitry at different levels of abstraction.} (A) A vertical section of chicken retina stained with fluorescent labels, revealing the layered structure: outer nuclear layer (ONL) with photoreceptor cells (PRs); outer plexiform layer (OPL) with synapses between PRs, horizontal cells (HCs), and bipolar cells (BCs); inner nuclear layer housing HCs, BCs, and amacrine cells (ACs); inner plexiform layer with synapses between BCs, ACs, and ganglion cells (GCs); ganglion cell layer (GCL) housing the GCs. Photo credit: Andy Fisher~\cite{fisher_vertical_2008}. (B) The five classes of retinal neurons, illustrating the range of shapes (schematized) within each class. From~\cite{masland_neuronal_2012}. (C) Synaptic connections among the five cell classes. Closed vs open circles: sign-preserving vs sign-inverting synapses. Zig-zag symbol: electrical synapses. (D) Schematic representation of the standard model (Eqns \ref{eqn:GC1} and \ref{eqn:GC2}). The light stimulus $S(\vec x, t)$ gets distributed into many parallel bipolar cell channels $B(\vec x, t)$, whose outputs get pooled into the ganglion cell response $G(\vec x, t)$. Right hand side includes explicit amacrine cell signals (Eqn \ref{eqn:GC2}). Inset defines each element as a cascade of a linear spatiotemporal filter $K(\vec x, t)$ and a nonlinearity $N()$.}
    \label{fig:circuits}
\end{figure}

\section{Biological ingredients of the Standard Model}

\subsection{The eye}
The standard model largely treats the eye as a camera that converts the external light field into an intensity pattern on the retina.
In many experiments that contributed to the standard model, the eye was removed and images were projected directly on the retina. 
Of course, for a full account of real-world vision, the details of ocular function matter. This includes the deviations from the ideal camera model, like chromatic aberration, axial aberrations, and stochastic eye movements. 

\subsection{Neurons}
There are about 100 types of neuron in the retina, with recent estimates ranging to 140~\cite{yan_mouse_2020,hahn_evolution_2023}. 
They are grouped into 5 classes by criteria of connectivity (Fig~\ref{fig:circuits}C): photoreceptors (PRs) get input from light; ganglion cells (GCs) send output through the optic nerve; bipolar cells (BCs) connect photoreceptors to ganglion cells; horizontal cells (HCs) communicate among photoreceptors and bipolar cells; amacrine cells (ACs) communicate among bipolar and ganglion cells.
These neurons are organized in 3 layers of cell bodies, separated by 2 layers of synapses (Fig~\ref{fig:circuits}A).
Within each class, one can distinguish subtypes by their shape (Fig~\ref{fig:circuits}B). 
For example, about 12 bipolar cell types differ by the precise lamination of their axons and dendrites in the two synaptic layers~\cite{wassle_cone_2009, euler_retinal_2014}. 
Increasingly the classic morphological distinctions among these subtypes are also supplemented by a molecular classification based on gene expression patterns~\cite{shekhar_comprehensive_2016,hahn_evolution_2023}.

Importantly, this classification of cell types is not merely a subjective choice ``in the eye of the beholder''
\footnote{...or in the output of a UMAP plot...}
, but supported by objective criteria derived from the organization of the retina. 
For example, neurons of the same type are spaced out within the plane of the retina so as to ``tile'' the space more or less uniformly. 
By correlating the 2-dimensional arrays of neurons in the retina, one can observe a type of ``repulsion'' among neurons of the same cell type (at least in some cases), but not across types~\cite{kozlowski_retinal_2024,wassle_dendritic_1981,rockhill_spatial_2000}. 
Another observation (in the mouse retina) is that every cone photoreceptor connects to 10 bipolar cells, of 10 different anatomical types~\cite{wassle_cone_2009}. 
These features of retinal organization reveal that there is biological ground truth to the cell types that have been identified.
\footnote{
The retina also contains a few glial cell types: Muller cells, which are specific to the retina, astrocytes, and microglia.
While these cells clearly play important roles for retinal biology, it is fair to say that they don't appear in the Standard Model.
}

\subsection{Connections}
Much as the standard model of particle physics involves particles and their interactions, here we must specify neurons and their connections.
The five major classes form chemical synapses, but only as allowed by the rules of Figure~\ref{fig:circuits}C. Along the straight-through pathway, photoreceptors can either excite or inhibit bipolar cells, and bipolar cells excite ganglion cells. 
\footnote{Here ``excite'' stands for sign-preserving, and ``inhibit'' for sign-inverting synaptic transmission.}
In addition, photoreceptors excite horizontal cells, which in turn inhibit photoreceptors, and may excite or inhibit bipolar cells. Similarly, bipolar cells excite amacrine cells, which in turn inhibit bipolar cells, other amacrine cells, and ganglion cells.
\footnote{The occasional odd neuron type violates these rules, for example cell with amacrine morphology but excitatory ribbon synapses like those of bipolar cells~\citep{della_santina_glutamatergic_2016,shekhar_comprehensive_2016}. 
This discovery came as a great surprise, underlining the remarkable reliability of the connectivity rules.}
In addition one finds electrical connections via gap junctions, but primarily among neurons of the same type. 
For example, horizontal cells and certain amacrine cells form meshworks by which signals are shared laterally~\cite{bloomfield_diverse_2009}.

On a fine scale, there are precise rules that regulate which types connect to each other. The ``inner plexiform layer'' (Fig~\ref{fig:circuits}A) is a remarkable switchboard, about 40 µm thick (in the mouse), free of cell bodies, and devoted entirely to the synapses between bipolar, amacrine, and ganglion cells. Here, particular connections are segregated into sublayers. For example, the dendritic trees of certain ganglion cells occupy a precise lamina in the IPL only 1 µm thick~\cite{bae_digital_2018,helmstaedter_connectomic_2013}. 
In that lamina they can make connections with only certain types of bipolar and amacrine cells~\cite{siegertGeneticAddressBook2009,masland_neuronal_2012,seung_neuronal_2014}. 
The specific partner relationships are governed in part by the expression of cell adhesion molecules on the surface of pre- and post-synaptic cell~\cite{sanes_synaptic_2020,stevens-sostre_cellular_2024}. 

\subsection{Plasticity}
The function of neurons and synapses in the retina can be modulated substantially by the history of their activity, the external stimuli, and other context, such as the time of day. 
Perhaps the largest effects are those of light adaptation in the photoreceptors, which can change the sensitivity of the light response by many log units. 
The biochemistry involved in retinal light adaptation is well-understood, and it can be
captured effectively using a mathematical model~\cite{fain_adaptation_2001,lamb_photoreceptor_2022,clark_dynamical_2013}. 
Many of the chemical synapses in the retina are subject to short-term facilitation and depression, owing to the history of pre- and post-synaptic activity~\cite{deng_short-term_2024,nikolaev_synaptic_2013}.
The conductance of gap junctions between like-type neurons is controlled by certain neuromodulators, and this can fundamentally alter the degree of lateral signal flow in the retina~\cite{roy_dopaminergic_2019}.
In this way the input-output relationship of the retina is not a static target, but may vary considerably on time scales from tenths of a second to a day.

\section{The response function of ganglion cells in the standard model}

On the basis of these ingredients, the standard model develops the input-output function of the retina, by which one can predict the response of a retinal ganglion cell based on the visual stimulus presented to the eye. 
The canonical form for this response function is the ``rectified subunit model''~\cite{enroth-cugell_receptive-field_1987,zapp_retinal_2022,victor_nonlinear_1979,hochstein_linear_1976,gollisch_modeling_2008,wienbar_dynamic_2018}. 
It postulates that the output of photoreceptors branches into multiple parallel pathways, which correspond to different bipolar cell types.
The ganglion cell, in turn, recombines input from these parallel pathways by summing over bipolar cell outputs.
Crucially, the output of each bipolar cell can be rectified, owing to the non-linearity of transfer at the bipolar cell synapse. 
Thus, each bipolar cell becomes a ``rectified subunit'' of the ganglion cell's visual receptive field.

Mathematically, one can formulate the model as follows. We start with the visually driven input experienced by a bipolar cell of type $i$ located at position $\vec x$ on the retina:

\begin{equation}
    B_{i}(\vec x, t) = \int_{\vec x', t'} K_i\left(\vec x' - \vec x, t' - t\right) S \left( \vec x', t' \right) {\rm{d}}^2x' {\rm{d}}t'
    \label{eqn:BC}
\end{equation}

where $S \left( \vec x', t' \right)$ is the light  stimulus at point $\vec x'$ and time $t'$. The response kernel $K_i\left(\vec x' - \vec x, t' - t\right)$ spells out how a bipolar cell of type $i$ located at position $\vec x$ on the retina weights the stimulus at point $\vec x'$ and time $t'$.

At the next stage, a ganglion cell of type $j$ pools the outputs of many bipolar cells:

\begin{equation}
    G_{j}(\vec x, t) = \sum_i \sum_{\vec x'} w_{ji}\left(\vec x' - \vec x\right) N_{ji}\left( B_{i}(\vec x', t)\right)
    \label{eqn:GC1}
\end{equation}

Here $w_{ji}\left(\vec x' - \vec x \right)$ is the weight by which a BC of type $i$ at location $\vec x'$ contributes to a GC of type $j$ at location $\vec x$. The function $N_{ji}()$ represents the non-linearity of transmission from BC type $i$ to GC type $j$.
Typically this is modeled as a simple rectifying function or a sigmoid~\cite{wienbar_dynamic_2018,zapp_retinal_2022}.

Where do the other neuronal classes fit into this scheme? Horizontal and amacrine cells contribute by shaping the responses of bipolar cells. For example, a typical bipolar cell kernel $K_i\left(\vec x' - \vec x, t' - t\right)$ includes an antagonistic surround region, which results from lateral signal flow in the horizontal and amacrine networks. Similarly, the transient kinetics of the typical bipolar cell kernel are shaped in part by negative feedback from amacrine cells onto the bipolar cell terminal. Many applications of the rectified subunit model leave it at that (Eqn \ref{eqn:GC1}), without an explicit modeling of amacrine or horizontal cells.

In other cases, it is necessary to include inhibition from amacrine cells directly~\cite{geffen_retinal_2007,baccus_retinal_2008,manookin_neural_2018,vaney_direction_2012}. Because amacrines also pool over bipolar cell inputs, one can model their response like that of ganglion cells,

\begin{equation}
    A_{k}(\vec x, t) = \sum_i \sum_{\vec x'} u_{ki}\left(\vec x' - \vec x\right) L_{ki}\left( B_{i}(\vec x', t)\right)
    \label{eqn:AC}
\end{equation}

and then add their output to the ganglion cell signal

\begin{equation}
    G_{j}(\vec x, t) = \sum_i \sum_{\vec x'} w_{ji}\left(\vec x' - \vec x\right) N_{ji}\left( B_{i}(\vec x', t)\right) + \sum_k \sum_{\vec x''} v_{jk}\left(\vec x'' - \vec x\right) M_{ji}\left( A_{k}(\vec x'', t)\right)     
    \label{eqn:GC2}
\end{equation}

Here $ u_{ki}$ and $L_{ki}\left(\right)$ are the weights and nonlinearities of synapses from BCs to ACs, and likewise $ v_{ki}$ and $M_{ki}\left(\right)$ for synapses from ACs to GCs.

Modeling using this full generality (Eqn \ref{eqn:GC2}) is rare. For example, many applications assume a simple common form for all the rectifying nonlinearities (e.g. $N_{ji}()$ in Eqn \ref{eqn:GC1}), but there are known occasions where a finer distinction is needed. For example the synaptic output of ON bipolar cells tends to be more linear than for OFF bipolars~\cite{turnerSynapticRectificationControls2016,dembBipolarCellsContribute2001}. 

\section{Core features and extensions of the standard model}

The response function of Eqn \ref{eqn:GC1} is essentially a multi-channel linear-nonlinear cascade model~\cite{hunter_identification_1986,schwartz_spike-triggered_2006,gollisch_modeling_2008} (Fig~\ref{fig:circuits}D). 
The common visual stimulus $S$ gets processed in parallel channels, each of which applies a linear spatio-temporal filter $K$. The output of each channel gets passed through a static nonlinear function $N$. The results get combined in a weighted sum to form the GC signal $G$. Optionally there is another nonlinearity that converts this signal into spikes. 

This general framework is constrained in a number of ways that reflect important aspects of retinal processing. Here I discuss the main features of the ``plain vanilla'' model (Eqn \ref{eqn:GC1}), and how they can be enhanced for special purposes.

\subsection{Translation-invariance}

The kernels $K_i$ and weighting functions $w_{ji}$ are at the heart of the model, because they specify the projection from light stimuli to BCs and from BCs to GCs. In the standard model, these weights depend on the position of inputs and outputs only through the difference vector, $\vec x' - \vec x$. As a result, the transformation from visual stimulus to spikes of a certain GC type follows the same rules at every point on the retina. 

In reality, the retina is translation-invariant only to a first approximation. For example, in the primate retina, both the size and density of GC receptive fields vary substantially from the fovea to the periphery~\cite{grunert_cell_2020}. Similar systematic gradients in response properties of a given cell type also exist in other species~\cite{bleckert_visual_2014,zhang_most_2012,yin_physiology_2009}. 
In practice, many functional experiments on retinal responses are performed over small regions of retina where the gradients don't come into play. In other cases, of course, one can expand on the standard model by adding some regional variation to the kernels.

\subsection{Linear filters}
The bipolar cell signal is commonly taken to be a linear function of the light stimulus, obtained by a simple convolution in space and time (Eqn \ref{eqn:BC}). 
This may seem surprising, given the strong nonlinearities that are apparent in the photoreceptor response, in particular the huge modulation of receptor gain depending on recent light exposure. 
In practice, many experiments are conducted under conditions that drive those nonlinearities only weakly, for example by modulating the light intensity over a limited range. 
However, natural vision includes stimuli with high local intensity fluctuations. In modeling retinal function under those conditions, it may be useful to include a nonlinear gain control leading to the bipolar cell signal~\cite{karamanlis_retinal_2022,yuAdaptationConePhotoreceptors2022}. 

\subsection{Dynamics} 
In its simplest formulation above (Eqn \ref{eqn:GC2}), the standard model places all the dynamics into the response kernel of the bipolar cell. None of the other elements introduces a time-dependence. 
This simplification reflects the fact that the photoreceptor is by far the slowest neuron in the retina and sets the basic time scale for retinal signals. By comparison, neurons downstream of the bipolar cell are relatively fast. 

For some purposes, however, it may be important to include additional dynamic components. For example, the input that the ganglion cell receives from ACs will be slightly delayed relative to the input from BCs, due to the additional synapse involved. As a result, this negative feed-forward loop can implement a high-pass filter that makes the ganglion cell exquisitely sensitive to change~\cite{nirenberg_light_1997,cafaro_regulation_2013,diamond_inhibitory_2017}. 
Another important dynamic sfeature may be short-term depression at bipolar or amacrine cell synapses, which has been invoked for certain adaptation phenomena~\cite{jarsky_synaptic_2011,wan_synaptic_2011,hosoya_dynamic_2005,ebert_temporal_2024}. 

\subsection{Nonlinearity} 
The standard model locates all the nonlinearity in the synaptic transfer from bipolar (or amacrine) to ganglion cell (Eqn \ref{eqn:GC2}). 
Without that nonlinear transmission, the ganglion cell response becomes a simple linear function of the stimulus. 
All the pooling over bipolar cell of different types and locations just leads to a simple average of their response kernels. 
Not much visual computation can be accomplished this way, beyond some spatial or temporal filtering. 

By contrast, a system of parallel pathways, each followed by a rectifying nonlinearity, has very rich computational power. In fact, such a model serves as a universal approximator: With the proper choice of kernel $K$ in each pathway one can in principle compute any desired function of space and time~\cite{cybenko_approximation_1989,hornik_multilayer_1989}.  

\subsection{Point neurons}
In the plain vanilla model, each of the interneurons (bipolar or amacrine cells) sums all its synaptic inputs and produces a single output signal, communicated through a nonlinear synapse. In other words the cells are treated as ``point neurons'', which greatly simplifies the mathematical treatment of the circuit.

However, there are known cases where different parts of the neuron carry different signals. For example the starburst amacrine cell -- a central element in the circuit for direction-selectivity -- 
seems to carry out different motion computations in each segment of its dendritic arbor~\cite{eulerDirectionallySelectiveCalcium2002,vaney_direction_2012}. 
From a modeling perspective, one can assign each of these dendrites to be another LN subunit, converting bipolar cell inputs to GABA-ergic output onto the ganglion cell.

Another example occurs at the synaptic terminals of bipolar cells, which often receive inhibitory input from amacrine cells directly adjacent to the bipolar cell's synaptic release site. 
These local interactions do not spread to the other terminals, and thus each terminal may compute its own function of the visual stimulus~\cite{asari_divergence_2012,matsumoto_direction_2021}. 
Again, for modeling purposes, one could treat each bipolar terminal as another LN subunit that combines input from bipolar and amacrine cell signals.

\subsection{Spiking}
As formulated above, neuronal responses are continuous functions of time. In fact, most neurons in the retina do not produce action potentials, except for the GCs themselves and some types of ACs\footnote{...and possibly BCs under certain conditions}. 
For the GCs, one can interpret the continuous time series $G(t)$ as an instantaneous firing rate, and compare it to the measured spike rate of a neuron, for example averaged over identical stimulus trials. 
Or one may interpret $G(t)$ as a membrane current and feed it through a spike-generating model to predict a ganglion cell spike train~\cite{keat_predicting_2001,pillow_spatio-temporal_2008}. 

\section{Why is the Standard Model a useful abstraction?}

As hinted above, the Standard Model (Eqn \ref{eqn:GC2}) can be seen as one point along a continuum of models for retinal function: ranging from the simplest, in which each ganglion cell just reports the amount of light in a circular receptive field
\footnote{Remarkably, this pixel detector view of the retina is still en vogue in many discussions of higher level vision.}
, to a detailed compartmental model of membranes, receptors, channels, and vesicle release sites. 
Along that continuum, the Standard Model is a uniquely powerful abstraction of biological reality, because
(1) it can explain a remarkable range of phenomena, compared to the limited complexity of its ingredients;
(2) it is mathematically simple enough that one can reason about its behavior even without explicit calculation; 
and (3) theorems have been proven about the expressivity of this model that hint at an evolutionary theory for the structure of the retina~\cite{cybenko_approximation_1989,hornik_multilayer_1989}. 

What does a retinal response model need to explain? Ideally, one would like to predict the spike train of a ganglion cell in response to any visual stimulus on the retina. Of course, the stimulus $S(x,t)$ has infinite dimensionality, so one can only probe that space sparsely, and choices need to be made. Early on, those choices were motivated by technical feasibility and mathematical convenience (e.g. flashing spots, or traveling gratings). More recently there has been a turn towards stimuli that actually happen ecologically (natural images and movies, eye movements)~\cite{qiu_natural_2021,brackbill_reconstruction_2020,karamanlis_nonlinear_2025,maheswaranathan_interpreting_2023}. In the process, a host of interesting visual computations have been identified across the $\sim$40 ganglion cell types~\cite{grunert_cell_2020,sanes_types_2015}. Many of these aspects of retinal responses seemed puzzling at first, but eventually were seen to follow from the same standard model, if one merely adjusts the parameters~\cite{schwartz_retinal_2021,gollisch_eye_2010,gollisch_features_2013,kastner_insights_2014,kerschensteiner_feature_2022}. A non-exhaustive list of these phenomena includes: 

\textbf{Center-surround antagonism:} A generic feature of responses in many ganglion cell types is the antagonism between the center of the receptive field and the surround region. 
For example, an ON cell is excited by light falling on the center but suppressed by light on the surround, leading to the classic ``mexican hat'' sensitivity profile. 
In the standard model, the bipolar cell subunits already have an antagonistic surround, and further surround inhibition comes from amacrine cell subunits. The combination can account for any receptive field profile via the weighting functions $w$ and $u$~\cite{enroth-cugell_receptive-field_1987,turner_receptive_2018}. 

\textbf{Non-linear spatial summation:} The proposal for a non-linear subunits model first emerged from a puzzling observation about GCs in the cat. So-called X-cells act as though they linearly sum the light in the receptive field, but Y-cells don't. They fire every time one redistributes the light within the receptive field, even while leaving the total unchanged~\cite{enroth-cugell_contrast_1966,hochstein_linear_1976,victor_nonlinear_1979,krieger_four_2017}. 
In the standard model, this can be explained by making the bipolar cell output functions ($N$ in Eqn~\ref{eqn:GC1}) more linear (X-type) or more rectifying (Y-type). In the Y-cell, any change in light pattern excites some subunits but inhibits others. Owing to the output rectification, only the excitation gets passed on to the ganglion cell and produces a burst of spikes. 
The nonlinear subunits in the model have since been identified as bipolar cells~\cite{dembBipolarCellsContribute2001,turnerSynapticRectificationControls2016}. 

\textbf{Texture sensitivity:} Ganglion cells often respond strongly to texture patterns that are much finer than the size of the receptive field. By localizing bipolar cells and mapping their non-linear output function, one can explain such texture responses in detail~\cite{schwartz_spatial_2012,gollisch_features_2013}. 

\textbf{Direction selectivity:} Some ganglion cells respond strongly when an object moves through the receptive field in one direction, but not in the opposite direction. 
In part, this is explained by a spatial offset between excitatory (BC) and delayed inhibitory (starburst AC) inputs to the ganglion cell~\cite{barlowMechanismDirectionallySelective1965,helmstaedter_connectomic_2013}, 
This can be captured by the synaptic weights $w$ and $u$ in (Eqn \ref{eqn:GC2}).
As mentioned above, another component of direction-selectivity relies on dendritic processing within the starburst cell~\cite{vaney_direction_2012}.

\textbf{Object motion sensitivity:} Some ganglion cells fire when a textured object within the receptive field moves differently than its background. If the entire scene moves across the retina in concert -- as happens during eye movements -- these GCs remain silent. A version of the standard model with the proper choice of nonlinearity and amacrine cell contribution (Eqn \ref{eqn:GC2}) was shown to match these response properties in detail~\cite{olveczky_segregation_2003,baccus_retinal_2008,jacoby_three_2017}. 

\textbf{Sensitivity to approaching motion:} Many retinal ganglion cells respond more strongly when a patterned surface approaches than when it recedes from the eye~\citep{munch_approach_2009,appleby_selectivity_2020}. 
This unexpected bias emerges naturally from the standard model, through the interplay between non-linear subunits in the center and the surround of the GC's receptive field~\cite{appleby_selectivity_2020}. 
The output nonlinearity for both types of subunits is an essential feature of this mechanism. 

\textbf{Motion anticipation:} Certain ganglion cells seem to ``anticipate'' the trajectory of a moving object. These neurons fire near the leading edge of the object on the retina, rather than behind the object, as one might expect based on slow photoreceptors and synaptic delays~\cite{berry_anticipation_1999,leonardo_nonlinear_2013}. 
There is no precognition involved. Rather one can explain the effects based on nonlinear inputs from BCs and ACs, and some simple assumptions about their spatial distribution on the GC dendritic field~\cite{johnston_general_2015}. 

\textbf{Spike-latency coding:} Some ganglion cells fire a short burst of spikes following both a brightening or a dimming within the receptive field. But the response latency is shorter to a dimming. In this way, the timing of the first spike following an image change on the retina can convey the content of the image~\cite{gollisch_rapid_2008}. This behavior results naturally because the GC sums over both on-type and off-type bipolar cells, and the kinetics ($K(t)$) of the On BCs are slightly slower~\cite{gollisch_modeling_2008}, presumably because of delays in the metabotropic glutamate receptor pathway. 


\section{Lessons from the Standard Model}
Why is it that we understand the retina so well compared to just about any other part of the central nervous system? 
As a neural circuit with over 100 recognized cell types, and employing most of the known neurotransmitters and modulators, the retina certainly does not lack in complexity. 
A number of fortuitous conditions helped enable the dissection of this system.

At the conceptual level, the retina is self-contained, and it has clearly defined input and output populations, such that the explanations of function can be found entirely within the retina.
An important technical advantage is that one can isolate the retina in a dish while maintaining most of its function, which makes its innards accessible to electrodes, pharmacology, and other tools.
Furthermore, one can stimulate the entire input layer at single-cell resolution simply by shining patterned light on the circuit.
This latter advantage can be ported to other systems through the advent of optogenetics. For example one can now stimulate the input neurons to the olfactory bulb with light patterns, which should assist in developing a quantitative model of neural processing in that circuit~\cite{chong_manipulating_2020}.

As a complement to the list of fortuitous conditions favoring retinal neuroscience, it is useful to consider what concepts and approaches were \emph{not needed} for success in this domain.

\subsection{Oscillations?}
When I started my first job, a senior professor from a rival university explained at the whiteboard that the retina is just an array of coupled oscillators. And if you only analyze it that way, you will understand how it performs Fourier transforms of the visual image. Shortly thereafter a physicist came to my office to explain how the major problem holding neuroscience back is that people don’t know how to compute power spectra properly. For a young scientist entering a new discipline this felt disconcerting. Is it really all about oscillations? By now, 30 years later, we can be fairly sure about two things: There is no sense in which the retina performs a Fourier transform on the image. And second, there is no evidence that oscillations and coupled oscillators are a useful way to think about retinal processing.

Sometime in the 1990s, a number of articles in high-profile journals claimed that in fact, spontaneous oscillations were an important part of visual encoding of moving objects. In the visual cortex of the cat, neurons representing the same object would oscillate at the same frequency, and those for another object at a different frequency. This was touted as a neural mechanism to solve the ``binding problem''~\cite{engel_interhemispheric_1991}. Subsequent articles traced the root of those oscillations ultimately all the way back to the retina~\cite{neuenschwander_long-range_1996}. This raised some consternation in the retinal research community
\footnote{...though not among the reviewers working for the high-impact magazines...}
because there had been hundreds of reports on retinal responses to moving objects, and very few of those mentioned any oscillations. Eventually these oscillations were found to be a side effect of anesthesia~\cite{neuenschwander_functional_2023}.
They don't occur in awake animals, and have nothing to do with the binding problem. 

All in all, the retina seems designed biologically to avoid ringing -- despite the combination of high gain and feedback loops that might present risk factors for runaway oscillation. As a result, there really is no role for oscillations in the Standard Model of the retina.
 
\subsection{Fine-grained causality?}
Much of neuroscience is concerned with explaining psychological functions based on neural mechanisms.
A predominant concern in this effort is the localization of function, such as ``Brodmann area 17 is for low-level processing of visual images''.
With the advent of modern cell typing, one now reads fine-grained claims like ``Cell type X in brain area Y implements behavioral function Z''.
There is little support for any such ``microphrenology'' in the retina.  

For example, in the psychology of vision, ``spatial contrast'' and ``color'' are two very different stimulus attributes. But their processing within retinal circuitry is entirely intertwined. Horizontal cells carry signals for lateral inhibition, which sharpens spatial contrast. But the same neurons are also essential in the processing of color~\cite{diamond_inhibitory_2017,martin_colour_1998,joesch_neuronal_2016}. 

On a local level within the circuit, the direction of causality isn't even well-defined. For example, the A2 amacrine cell drives cone ON-bipolar cells under dim light conditions, but gets driven by those neurons under bright conditions~\cite{oesch_illuminating_2011}.
The prominence of feedback loops throughout the retina further complicates any assessment of which neurons lie upstream vs downstream.

In summary, the retinal circuit simply functions as a whole, and one cannot usefully assign sub-functions to specific cell types. This may prove difficult in other brain regions as well. 

\subsection{Detailed compartmental modeling?}
There is some enthusiasm in current neuroscience for treating the single neuron as a powerful processor in its own right, whose functions can only be understood through detailed models with hundreds of compartments harboring local nonlinearities~\cite{london_dendritic_2005}. The Standard Model largely treats retinal cells as single-compartment point neurons, meaning that a single activity variable is sufficient to describe the neuron's dynamics. This is not to say that all retinal neurons are electrotonically compact, in fact there is much evidence for local signaling within the axonal trees of bipolar cells, or the dendritic trees of amacrine and ganglion cells that is not shared with other parts of the neuron~\cite{hartveit_dendritic_2022,vaney_direction_2012,matsumoto_direction_2021}. 

Nonetheless, the point-neuron approximation made in the Standard Model accounts for most phenomena of interest. Take, for example, the most intensely debated example of retinal computation: direction-selectivity. Yes, there is evidence for spikes in the dendrites of direction-selective ganglion cells. Also, there may be a contribution from spatially ordered excitation and inhibition in the dendrites. Nonetheless, a large contributor to direction-selectivity is that the ganglion cell gets inhibition from starburst amacrines located on one side of its dendritic tree but not on the other~\cite{vaney_direction_2012}. In practice one can understand the phenomenon as arising from a spatially asymmetric circuit between point neurons, much as envisioned 60 years ago by the discoverers~\citep{barlowMechanismDirectionallySelective1965}. The lesson here is that, while single neurons may harbor prodigious computational capacity, that may not always be the way nature uses them. When faced with brain circuits whose function remains mysterious it may still pay off to start with simple point-neuron models.

\subsection{Low-dimensional neural dynamics?}
Another popular meme in current neuroscience is that underlying the responses of millions of neurons are just a few latent variables. After recording thousands of signals, one should therefore look to project them onto some low-dimensional manifold in which the truly important neural variables are represented~\citep{cunningham_dimensionality_2014}. Typically there are just two or three of these, which is fortuitously convenient, because we can then draw those manifolds on paper. 

In the case of the retina, this approach is obviously wrong. The human retina has a million neurons as outputs, and retinal circuitry is organized so that those million outputs report different things. For example, the receptive fields of neurons of the same type overlap only minimally, so they avoid redundant signaling. To good approximation, at every point in space there exists just one neuron of any given cell type. So there is no low-dimensional latent manifold hidden inside the million dimensions of the retinal output. The same argument holds for the early stages of visual cortex, where the retinotopic map alone ensures that different neurons encode different things. 

The lesson for other areas of the brain is that we may simply be ignorant about all the items represented there. It seems somehow implausible that the primary visual cortex employs a billion neurons to process the 1-million-dimensional optic nerve input, whereas the pre-frontal cortex uses a billion neurons to represent the dynamics of just 3 variables. Obviously, any single behavioral experiment that only involves a handful of variables cannot bring out the richness of the neural representation. We would not have arrived at the present understanding of retinal coding by using exclusively low-dimensional stimuli.

\section{Theories of the retina}
A major remaining remaining challenge is to understand where the Standard Model comes from. By analogy to particle physics, we now know what all the particles are, namely the cell types. We also know about the forces between particles: the synaptic interactions. Also we believe that all phenomenology of retinal function ultimately traces back to those cell types and their interactions. But we don't have a theory of why the cell types are what they are. 

In the case of particle physics, there may be no ultimate explanation: The present pantheon of particles may just boil down to a random parameter setting in the initial Big Bang. In the case of retinal structure and function, we expect that there is a deeper explanation related to evolution, with natural selection favoring one type of structure over others.

\subsection{Efficient coding vs selective computation}
Today, the dominant normative framework for the retina is efficient coding theory~\cite{barlowPossiblePrinciplesUnderlying1961,simoncelli_vision_2003,manookin_two_2023}. It postulates that the optic nerve is a bottleneck for visual information, and the retina reformats the raw image from photoreceptors so it can pass through the bottleneck. This involves removing redundancy in the visual signal that results from the dominance of low frequencies (both spatial and temporal) within natural scenes. 

This theory has been very successful at explaining some qualitative features of retinal processing: lateral inhibition in space, temporal band-pass filtering, and the split into On and Off pathways. 
But it fails to account for the large diversity of cell types among retinal ganglion cells. Redundancy reduction really recommends just two GC types -- one On one Off -- with a spatio-temporal receptive field adapted to the natural scene statistics.
Instead we have about 40 GC types. How to explain this discrepancy?

On a larger scale, one might say that the visual system is more concerned with discarding information than preserving it. Our perceptual system ultimately extracts only about 10 bits per second from the visual image, whereas the capacity of the cone photoreceptors is about 10$^9$ bits/s~\citep{zheng_unbearable_2025}. 
Whittling the information down by 8 log units is a massive challenge, and it seems likely that the retina takes a first cut in discarding information that will not ultimately be used. 
Thus it may be more useful to treat retinal processing as selective computation with loss of information, rather than focusing on efficient coding.

\subsection{Why so many ganglion cell types?}
Still this leaves open the puzzle of the many cell types. 
One attractive hypothesis here is that each ganglion cell type precomputes a certain visual feature that can be used directly for a specific behavioral need. For example, a molecularly identified type of ganglion cell in the mouse responds to slow image motion on the retina. These neurons project specifically to the accessory optic system, a set of nuclei implicated in measuring retinal slip of the visual image and controlling eye movements to compensate~\citep{yonehara_identification_2009}. This neuron type appears to solve a special function to support active vision.

In other cases, however, the connection between visual behaviors and retinal function is more complicated. For example, the looming reaction of the mouse is a very robust visual behavior, by which the animal escapes from an approaching aerial predator~\citep{yilmaz_rapid_2013}. Certain ganglion cells in the mouse retina that are sensitive to dark expanding objects~\citep{munch_approach_2009,kim_dendritic_2020,wang_off-transient_2021}. But those neurons also respond to many other stimuli, like flashed objects, unlike the behavioral looming reaction~\citep{krieger_four_2017}. 
Furthermore, signals from different GC types hardly remain separate, but get combined already via synaptic convergence in the thalamus and superior colliculus~\cite{reinhard_projection_2019}. Thus, the observed precision and selectivity of the looming reaction and other visual behaviors likely relies on the downstream combination of multiple retinal channels~\cite{hoy_defined_2019}. 

In summary, a useful theory of the retina will need to explain the great diversity of the many output channels and also the specific stimulus features that they represent. If successful, this will illuminate the principles of neural processing in downstream visual areas as well.

\newpage

\section*{Funding}
MM was supported by grants from the Simons Collaboration on the Global Brain (543015) and NIH (R01 NS111477).

\section*{Declaration of Interests}
The author declares no competing interests.

\section*{References}
\renewcommand{\bibsection}{}\vspace{0em}
\bibliographystyle{myunsrtnat}
\bibliography{references} 

\begin{thebibliography}{103}
\providecommand{\natexlab}[1]{#1}
\providecommand{\url}[1]{\texttt{#1}}
\expandafter\ifx\csname urlstyle\endcsname\relax
  \providecommand{\doi}[1]{doi: #1}\else
  \providecommand{\doi}{doi: \begingroup \urlstyle{rm}\Url}\fi

\bibitem[Kolb et~al.()Kolb, Nelson, Fernandez, and Jones]{kolb_webvision_nodate}
Kolb, H., Nelson, R., Fernandez, E., and Jones, editors.
\newblock \emph{Webvision – {The} {Organization} of the {Retina} and {Visual} {System}}.
\newblock URL \url{https://www.webvision.pitt.edu/}.

\bibitem[Schwartz(2021)]{schwartz_retinal_2021}
Schwartz, G.
\newblock \emph{Retinal {Computation}}.
\newblock Academic Press, London, 1st edition edition, August 2021.
\newblock ISBN 978-0-12-819896-4.

\bibitem[Cohen-Tannoudji et~al.(1977)Cohen-Tannoudji, Diu, and Laloë]{Cohen-Tannoudji:101367}
Cohen-Tannoudji, C., Diu, B., and Laloë, F.
\newblock \emph{Quantum mechanics; 1st ed.}
\newblock Wiley, New York, NY, 1977.
\newblock URL \url{https://cds.cern.ch/record/101367}.

\bibitem[Hahn et~al.(2023)Hahn, Monavarfeshani, Qiao, Kao, Kölsch, Kumar, Kunze, Rasys, Richardson, Wekselblatt, Baier, Lucas, Li, Meister, Trachtenberg, Yan, Peng, Sanes, and Shekhar]{hahn_evolution_2023}
Hahn, J., Monavarfeshani, A., Qiao, M., Kao, A.~H., Kölsch, Y., Kumar, A., Kunze, V.~P., Rasys, A.~M., Richardson, R., Wekselblatt, J.~B., Baier, H., Lucas, R.~J., Li, W., Meister, M., Trachtenberg, J.~T., Yan, W., Peng, Y.-R., Sanes, J.~R., and Shekhar, K.
\newblock Evolution of neuronal cell classes and types in the vertebrate retina.
\newblock \emph{Nature}, 624\penalty0 (7991):\penalty0 415--424, December 2023.
\newblock ISSN 1476-4687.
\newblock \doi{10.1038/s41586-023-06638-9}.

\bibitem[Fisher(2008)]{fisher_vertical_2008}
Fisher, A.
\newblock Vertical section of a chick retina {\textbar} 2008 {Photomicrography} {Competition}, 2008.
\newblock URL \url{https://www.nikonsmallworld.com/galleries/2008-photomicrography-competition/vertical-section-of-a-chick-retina}.

\bibitem[Masland(2012)]{masland_neuronal_2012}
Masland, R.~H.
\newblock The neuronal organization of the retina.
\newblock \emph{Neuron}, 76\penalty0 (2):\penalty0 266--280, October 2012.
\newblock ISSN 1097-4199.
\newblock \doi{10.1016/j.neuron.2012.10.002}.

\bibitem[Yan et~al.(2020)Yan, Laboulaye, Tran, Whitney, Benhar, and Sanes]{yan_mouse_2020}
Yan, W., Laboulaye, M.~A., Tran, N.~M., Whitney, I.~E., Benhar, I., and Sanes, J.~R.
\newblock Mouse {Retinal} {Cell} {Atlas}: {Molecular} {Identification} of over {Sixty} {Amacrine} {Cell} {Types}.
\newblock \emph{Journal of Neuroscience}, 40\penalty0 (27):\penalty0 5177--5195, July 2020.
\newblock ISSN 0270-6474, 1529-2401.
\newblock \doi{10.1523/JNEUROSCI.0471-20.2020}.
\newblock URL \url{https://www.jneurosci.org/content/40/27/5177}.
\newblock Publisher: Society for Neuroscience Section: Research Articles.

\bibitem[Wässle et~al.(2009)Wässle, Puller, Muller, and Haverkamp]{wassle_cone_2009}
Wässle, H., Puller, C., Muller, F., and Haverkamp, S.
\newblock Cone contacts, mosaics, and territories of bipolar cells in the mouse retina.
\newblock \emph{J Neurosci}, 29:\penalty0 106--17, January 2009.
\newblock \doi{10.1523/JNEUROSCI.4442-08.2009}.

\bibitem[Euler et~al.(2014)Euler, Haverkamp, Schubert, and Baden]{euler_retinal_2014}
Euler, T., Haverkamp, S., Schubert, T., and Baden, T.
\newblock Retinal bipolar cells: elementary building blocks of vision.
\newblock \emph{Nat Rev Neurosci}, 15:\penalty0 507--19, August 2014.

\bibitem[Shekhar et~al.(2016)Shekhar, Lapan, Whitney, Tran, Macosko, Kowalczyk, Adiconis, Levin, Nemesh, Goldman, McCarroll, Cepko, Regev, and Sanes]{shekhar_comprehensive_2016}
Shekhar, K., Lapan, S.~W., Whitney, I.~E., Tran, N.~M., Macosko, E.~Z., Kowalczyk, M., Adiconis, X., Levin, J.~Z., Nemesh, J., Goldman, M., McCarroll, S.~A., Cepko, C.~L., Regev, A., and Sanes, J.~R.
\newblock Comprehensive {Classification} of {Retinal} {Bipolar} {Neurons} by {Single}-{Cell} {Transcriptomics}.
\newblock \emph{Cell}, 166:\penalty0 1308--1323.e30, August 2016.
\newblock ISSN 0092-8674.
\newblock \doi{10.1016/j.cell.2016.07.054}.

\bibitem[Kozlowski et~al.(2024)Kozlowski, Hadyniak, and Kay]{kozlowski_retinal_2024}
Kozlowski, C., Hadyniak, S.~E., and Kay, J.~N.
\newblock Retinal neurons establish mosaic patterning by excluding homotypic somata from their dendritic territories.
\newblock \emph{Cell Reports}, 43\penalty0 (8):\penalty0 114615, August 2024.
\newblock ISSN 2211-1247.
\newblock \doi{10.1016/j.celrep.2024.114615}.
\newblock URL \url{https://www.sciencedirect.com/science/article/pii/S2211124724009653}.

\bibitem[Wässle et~al.(1981)Wässle, Peichl, and Boycott]{wassle_dendritic_1981}
Wässle, H., Peichl, L., and Boycott, B.~B.
\newblock Dendritic territories of cat retinal ganglion cells.
\newblock \emph{Nature}, 292:\penalty0 344--5, July 1981.

\bibitem[Rockhill et~al.(2000)Rockhill, Euler, and Masland]{rockhill_spatial_2000}
Rockhill, R.~L., Euler, T., and Masland, R.~H.
\newblock Spatial order within but not between types of retinal neurons.
\newblock \emph{Proc Natl Acad Sci U S A}, 97:\penalty0 2303--7, February 2000.
\newblock \doi{10.1073/pnas.030413497}.

\bibitem[Della~Santina et~al.(2016)Della~Santina, Kuo, Yoshimatsu, Okawa, Suzuki, Hoon, Tsuboyama, Rieke, and Wong]{della_santina_glutamatergic_2016}
Della~Santina, L., Kuo, S.~P., Yoshimatsu, T., Okawa, H., Suzuki, S.~C., Hoon, M., Tsuboyama, K., Rieke, F., and Wong, R. O.~L.
\newblock Glutamatergic {Monopolar} {Interneurons} {Provide} a {Novel} {Pathway} of {Excitation} in the {Mouse} {Retina}.
\newblock \emph{Current Biology}, 26\penalty0 (15):\penalty0 2070--2077, August 2016.
\newblock ISSN 0960-9822.
\newblock \doi{10.1016/j.cub.2016.06.016}.
\newblock URL \url{https://www.sciencedirect.com/science/article/pii/S0960982216306558}.

\bibitem[Bloomfield and Völgyi(2009)]{bloomfield_diverse_2009}
Bloomfield, S. and Völgyi, B.
\newblock The diverse functional roles and regulation of neuronal gap junctions in the retina.
\newblock \emph{Nature Reviews Neuroscience}, 10\penalty0 (7):\penalty0 495--506, July 2009.
\newblock ISSN 1471-003X.
\newblock \doi{10.1038/nrn2636}.

\bibitem[Bae et~al.(2018)Bae, Mu, Kim, Turner, Tartavull, Kemnitz, Jordan, Norton, Silversmith, Prentki, Sorek, David, Jones, Bland, Sterling, Park, Briggman, and Seung]{bae_digital_2018}
Bae, J.~A., Mu, S., Kim, J.~S., Turner, N.~L., Tartavull, I., Kemnitz, N., Jordan, C.~S., Norton, A.~D., Silversmith, W.~M., Prentki, R., Sorek, M., David, C., Jones, D.~L., Bland, D., Sterling, A. L.~R., Park, J., Briggman, K.~L., and Seung, H.~S.
\newblock Digital {Museum} of {Retinal} {Ganglion} {Cells} with {Dense} {Anatomy} and {Physiology}.
\newblock \emph{Cell}, 173\penalty0 (5):\penalty0 1293--1306.e19, May 2018.
\newblock ISSN 0092-8674.
\newblock \doi{10.1016/j.cell.2018.04.040}.
\newblock URL \url{https://www.sciencedirect.com/science/article/pii/S0092867418305725}.

\bibitem[Helmstaedter et~al.(2013)Helmstaedter, Briggman, Turaga, Jain, Seung, and Denk]{helmstaedter_connectomic_2013}
Helmstaedter, M., Briggman, K.~L., Turaga, S.~C., Jain, V., Seung, H.~S., and Denk, W.
\newblock Connectomic reconstruction of the inner plexiform layer in the mouse retina.
\newblock \emph{Nature}, 500:\penalty0 168--74, August 2013.
\newblock \doi{10.1038/nature12346}.

\bibitem[Siegert et~al.(2009)Siegert, Scherf, Del~Punta, Didkovsky, Heintz, and Roska]{siegertGeneticAddressBook2009}
Siegert, S., Scherf, B.~G., Del~Punta, K., Didkovsky, N., Heintz, N., and Roska, B.
\newblock Genetic address book for retinal cell types.
\newblock \emph{Nature Neuroscience}, 12\penalty0 (9):\penalty0 1197--1204, September 2009.
\newblock ISSN 1546-1726.
\newblock \doi{10.1038/nn.2370}.

\bibitem[Seung and Sumbul(2014)]{seung_neuronal_2014}
Seung, H.~S. and Sumbul, U.
\newblock Neuronal cell types and connectivity: lessons from the retina.
\newblock \emph{Neuron}, 83\penalty0 (6):\penalty0 1262--72, September 2014.
\newblock ISSN 0896-6273, 1097-4199.
\newblock \doi{10.1016/j.neuron.2014.08.054}.
\newblock Num Pages: 11 Place: Cambridge Publisher: Cell Press Web of Science ID: WOS:000342502400013.

\bibitem[Sanes and Zipursky(2020)]{sanes_synaptic_2020}
Sanes, J. and Zipursky, S.
\newblock Synaptic {Specificity}, {Recognition} {Molecules}, and {Assembly} of {Neural} {Circuits}.
\newblock \emph{CELL}, 181\penalty0 (3):\penalty0 536--556, April 2020.
\newblock ISSN 0092-8674.
\newblock \doi{10.1016/j.cell.2020.04.008}.

\bibitem[Stevens-Sostre and Hoon(2024)]{stevens-sostre_cellular_2024}
Stevens-Sostre, W.~A. and Hoon, M.
\newblock Cellular and {Molecular} {Mechanisms} {Regulating} {Retinal} {Synapse} {Development}.
\newblock \emph{Annual review of vision science}, 10\penalty0 (1):\penalty0 377--402, September 2024.
\newblock ISSN 2374-4642.
\newblock \doi{10.1146/annurev-vision-102122-105721}.
\newblock URL \url{https://pmc.ncbi.nlm.nih.gov/articles/PMC12022667/}.

\bibitem[Fain et~al.(2001)Fain, Matthews, Cornwall, and Koutalos]{fain_adaptation_2001}
Fain, G.~L., Matthews, H.~R., Cornwall, M.~C., and Koutalos, Y.
\newblock Adaptation in vertebrate photoreceptors.
\newblock \emph{Physiol Rev}, 81:\penalty0 117--151., 2001.

\bibitem[Lamb(2022)]{lamb_photoreceptor_2022}
Lamb, T.~D.
\newblock Photoreceptor physiology and evolution: cellular and molecular basis of rod and cone phototransduction.
\newblock \emph{Journal of Physiology-London}, 600\penalty0 (21):\penalty0 4585--4601, November 2022.
\newblock ISSN 0022-3751, 1469-7793.
\newblock \doi{10.1113/JP282058}.
\newblock Num Pages: 17 Place: Hoboken Publisher: Wiley Web of Science ID: WOS:000788266700001.

\bibitem[Clark et~al.(2013)Clark, Benichou, Meister, and Azeredo~da Silveira]{clark_dynamical_2013}
Clark, D.~A., Benichou, R., Meister, M., and Azeredo~da Silveira, R.
\newblock Dynamical adaptation in photoreceptors.
\newblock \emph{PLoS Comput Biol}, 9:\penalty0 e1003289, November 2013.
\newblock \doi{10.1371/journal.pcbi.1003289}.

\bibitem[Deng et~al.(2024)Deng, Oosterboer, and Wei]{deng_short-term_2024}
Deng, Z., Oosterboer, S., and Wei, W.
\newblock Short-term plasticity and context-dependent circuit function: {Insights} from retinal circuitry.
\newblock \emph{Science Advances}, 10\penalty0 (38):\penalty0 eadp5229, September 2024.
\newblock ISSN 2375-2548.
\newblock \doi{10.1126/sciadv.adp5229}.
\newblock Num Pages: 7 Place: Washington Publisher: Amer Assoc Advancement Science Web of Science ID: WOS:001317192900012.

\bibitem[Nikolaev et~al.(2013)Nikolaev, Leung, Odermatt, and Lagnado]{nikolaev_synaptic_2013}
Nikolaev, A., Leung, K.~M., Odermatt, B., and Lagnado, L.
\newblock Synaptic mechanisms of adaptation and sensitization in the retina.
\newblock \emph{Nat Neurosci}, 16:\penalty0 934--41, July 2013.
\newblock \doi{10.1038/nn.3408}.

\bibitem[Roy and Field(2019)]{roy_dopaminergic_2019}
Roy, S. and Field, G.~D.
\newblock Dopaminergic modulation of retinal processing from starlight to sunlight.
\newblock \emph{Journal of Pharmacological Sciences}, 140\penalty0 (1):\penalty0 86--93, May 2019.
\newblock ISSN 1347-8613, 1347-8648.
\newblock \doi{10.1016/j.jphs.2019.03.006}.
\newblock Num Pages: 8 Place: Kyoto Publisher: Japanese Pharmacological Soc Web of Science ID: WOS:000471882900013.

\bibitem[Enroth-Cugell and Freeman(1987)]{enroth-cugell_receptive-field_1987}
Enroth-Cugell, C. and Freeman, A.~W.
\newblock The receptive-field spatial structure of cat retinal {Y} cells.
\newblock \emph{J Physiol}, 384:\penalty0 49--79., 1987.

\bibitem[Zapp et~al.(2022)Zapp, Nitsche, and Gollisch]{zapp_retinal_2022}
Zapp, S.~J., Nitsche, S., and Gollisch, T.
\newblock Retinal receptive-field substructure: scaffolding for coding and computation.
\newblock \emph{Trends in Neurosciences}, 45\penalty0 (6):\penalty0 430--445, June 2022.
\newblock ISSN 0166-2236.
\newblock \doi{10.1016/j.tins.2022.03.005}.
\newblock URL \url{https://www.sciencedirect.com/science/article/pii/S016622362200056X}.

\bibitem[Victor and Shapley(1979)]{victor_nonlinear_1979}
Victor, J.~D. and Shapley, R.~M.
\newblock The nonlinear pathway of {Y} ganglion cells in the cat retina.
\newblock \emph{J Gen Physiol}, 74:\penalty0 671--89, December 1979.

\bibitem[Hochstein and Shapley(1976)]{hochstein_linear_1976}
Hochstein, S. and Shapley, R.~M.
\newblock Linear and nonlinear spatial subunits in {Y} cat retinal ganglion cells.
\newblock \emph{Journal of Physiology}, 262:\penalty0 265--84, 1976.

\bibitem[Gollisch and Meister(2008{\natexlab{a}})]{gollisch_modeling_2008}
Gollisch, T. and Meister, M.
\newblock Modeling convergent {ON} and {OFF} pathways in the early visual system.
\newblock \emph{Biol Cybern}, 99:\penalty0 263--78, November 2008{\natexlab{a}}.
\newblock \doi{10.1007/s00422-008-0252-y}.

\bibitem[Wienbar and Schwartz(2018)]{wienbar_dynamic_2018}
Wienbar, S. and Schwartz, G.~W.
\newblock The dynamic receptive fields of retinal ganglion cells.
\newblock \emph{Progress in Retinal and Eye Research}, 67:\penalty0 102--117, November 2018.
\newblock ISSN 1350-9462.
\newblock \doi{10.1016/j.preteyeres.2018.06.003}.
\newblock URL \url{https://www.sciencedirect.com/science/article/pii/S1350946218300284}.

\bibitem[Geffen et~al.(2007)Geffen, de~Vries, and Meister]{geffen_retinal_2007}
Geffen, M.~N., de~Vries, S.~E., and Meister, M.
\newblock Retinal ganglion cells can rapidly change polarity from {Off} to {On}.
\newblock \emph{PLoS Biol}, 5:\penalty0 e65, March 2007.
\newblock \doi{10.1371/journal.pbio.0050065}.

\bibitem[Baccus et~al.(2008)Baccus, Ölveczky, Manu, and Meister]{baccus_retinal_2008}
Baccus, S.~A., Ölveczky, B.~P., Manu, M., and Meister, M.
\newblock A retinal circuit that computes object motion.
\newblock \emph{J Neurosci}, 28:\penalty0 6807--17, July 2008.
\newblock \doi{10.1523/JNEUROSCI.4206-07.2008}.

\bibitem[Manookin et~al.(2018)Manookin, Patterson, and Linehan]{manookin_neural_2018}
Manookin, M.~B., Patterson, S.~S., and Linehan, C.~M.
\newblock Neural {Mechanisms} {Mediating} {Motion} {Sensitivity} in {Parasol} {Ganglion} {Cells} of the {Primate} {Retina}.
\newblock \emph{Neuron}, 97\penalty0 (6):\penalty0 1327--1340.e4, March 2018.
\newblock ISSN 0896-6273.
\newblock \doi{10.1016/j.neuron.2018.02.006}.
\newblock URL \url{https://www.cell.com/neuron/abstract/S0896-6273(18)30105-3}.
\newblock Publisher: Elsevier.

\bibitem[Vaney et~al.(2012)Vaney, Sivyer, and Taylor]{vaney_direction_2012}
Vaney, D.~I., Sivyer, B., and Taylor, W.~R.
\newblock Direction selectivity in the retina: symmetry and asymmetry in structure and function.
\newblock \emph{Nat Rev Neurosci}, 13:\penalty0 194--208, March 2012.
\newblock \doi{10.1038/nrn3165}.

\bibitem[Turner and Rieke(2016)]{turnerSynapticRectificationControls2016}
Turner, M.~H. and Rieke, F.
\newblock Synaptic {{Rectification Controls Nonlinear Spatial Integration}} of {{Natural Visual Inputs}}.
\newblock \emph{NEURON}, 90\penalty0 (6):\penalty0 1257--1271, June 2016.
\newblock ISSN 0896-6273.
\newblock \doi{10.1016/j.neuron.2016.05.006}.

\bibitem[Demb et~al.(2001)Demb, Zaghloul, Haarsma, and Sterling]{dembBipolarCellsContribute2001}
Demb, J.~B., Zaghloul, K., Haarsma, L., and Sterling, P.
\newblock Bipolar cells contribute to nonlinear spatial summation in the brisk-transient ({{Y}}) ganglion cell in mammalian retina.
\newblock \emph{J Neurosci}, 21:\penalty0 7447--54, October 2001.

\bibitem[Hunter and Korenberg(1986)]{hunter_identification_1986}
Hunter, I.~W. and Korenberg, M.~J.
\newblock The identification of nonlinear biological systems: {Wiener} and {Hammerstein} cascade models.
\newblock \emph{Biol. Cybern.}, 55:\penalty0 135--144, 1986.

\bibitem[Schwartz et~al.(2006)Schwartz, Pillow, Rust, and Simoncelli]{schwartz_spike-triggered_2006}
Schwartz, O., Pillow, J.~W., Rust, N.~C., and Simoncelli, E.~P.
\newblock Spike-triggered neural characterization.
\newblock \emph{Journal of Vision}, 6\penalty0 (4):\penalty0 13, July 2006.
\newblock ISSN 1534-7362.
\newblock \doi{10.1167/6.4.13}.
\newblock URL \url{https://doi.org/10.1167/6.4.13}.

\bibitem[Grünert and Martin(2020)]{grunert_cell_2020}
Grünert, U. and Martin, P.~R.
\newblock Cell types and cell circuits in human and non-human primate retina.
\newblock \emph{Progress in Retinal and Eye Research}, 78:\penalty0 100844, September 2020.
\newblock ISSN 1350-9462.
\newblock \doi{10.1016/j.preteyeres.2020.100844}.
\newblock URL \url{https://www.sciencedirect.com/science/article/pii/S1350946220300161}.

\bibitem[Bleckert et~al.(2014)Bleckert, Schwartz, Turner, Rieke, and Wong]{bleckert_visual_2014}
Bleckert, A., Schwartz, G.~W., Turner, M.~H., Rieke, F., and Wong, R.~O.
\newblock Visual space is represented by nonmatching topographies of distinct mouse retinal ganglion cell types.
\newblock \emph{Curr Biol}, 24:\penalty0 310--5, February 2014.
\newblock \doi{10.1016/j.cub.2013.12.020}.

\bibitem[Zhang et~al.(2012)Zhang, Kim, Sanes, and Meister]{zhang_most_2012}
Zhang, Y., Kim, I.~J., Sanes, J.~R., and Meister, M.
\newblock The most numerous ganglion cell type of the mouse retina is a selective feature detector.
\newblock \emph{Proc Natl Acad Sci U S A}, 109:\penalty0 E2391--8, September 2012.
\newblock \doi{10.1073/pnas.1211547109}.

\bibitem[Yin et~al.(2009)Yin, Smith, Sterling, and Brainard]{yin_physiology_2009}
Yin, L., Smith, R.~G., Sterling, P., and Brainard, D.~H.
\newblock Physiology and morphology of color-opponent ganglion cells in a retina expressing a dual gradient of {S} and {M} opsins.
\newblock \emph{J Neurosci}, 29\penalty0 (9):\penalty0 2706--24, March 2009.
\newblock ISSN 0270-6474.
\newblock \doi{10.1523/JNEUROSCI.5471-08.2009}.
\newblock Num Pages: 19 Place: Washington Publisher: Soc Neuroscience Web of Science ID: WOS:000263886100006.

\bibitem[Karamanlis et~al.(2022)Karamanlis, Schreyer, and Gollisch]{karamanlis_retinal_2022}
Karamanlis, D., Schreyer, H.~M., and Gollisch, T.
\newblock Retinal {Encoding} of {Natural} {Scenes}.
\newblock \emph{Annual Review of Vision Science}, 8:\penalty0 171--193, 2022.
\newblock ISSN 2374-4642, 2374-4650.
\newblock \doi{10.1146/annurev-vision-100820-114239}.
\newblock Num Pages: 23 Place: Palo Alto Publisher: Annual Reviews Web of Science ID: WOS:000856252600007.

\bibitem[Yu et~al.(2022)Yu, Turner, Baudin, and Rieke]{yuAdaptationConePhotoreceptors2022}
Yu, Z., Turner, M.~H., Baudin, J., and Rieke, F.
\newblock Adaptation in cone photoreceptors contributes to an unexpected insensitivity of primate {{On}} parasol retinal ganglion cells to spatial structure in natural images.
\newblock \emph{eLife}, 11:\penalty0 e70611, March 2022.
\newblock ISSN 2050-084X.
\newblock \doi{10.7554/eLife.70611}.

\bibitem[Nirenberg and Meister(1997)]{nirenberg_light_1997}
Nirenberg, S. and Meister, M.
\newblock The light response of retinal ganglion cells is truncated by a displaced amacrine circuit.
\newblock \emph{Neuron}, 18:\penalty0 637--50, April 1997.

\bibitem[Cafaro and Rieke(2013)]{cafaro_regulation_2013}
Cafaro, J. and Rieke, F.
\newblock Regulation of spatial selectivity by crossover inhibition.
\newblock \emph{The Journal of Neuroscience: The Official Journal of the Society for Neuroscience}, 33\penalty0 (15):\penalty0 6310--6320, April 2013.
\newblock ISSN 1529-2401.
\newblock \doi{10.1523/JNEUROSCI.4964-12.2013}.

\bibitem[Diamond(2017)]{diamond_inhibitory_2017}
Diamond, J.~S.
\newblock Inhibitory {Interneurons} in the {Retina}: {Types}, {Circuitry}, and {Function}.
\newblock In Movshon, J.~A. and Wandell, B.~A., editors, \emph{Annual {Review} of {Vision} {Science}, {Vol} 3}, volume~3, pages 1--24. Annual Reviews, Palo Alto, 2017.
\newblock ISBN 978-0-8243-5103-8.
\newblock \doi{10.1146/annurev-vision-102016-061345}.
\newblock ISSN: 2374-4642, 2374-4650 Num Pages: 24 Series Title: Annual Review of Vision Science Web of Science ID: WOS:000411715700001.

\bibitem[Jarsky et~al.(2011)Jarsky, Cembrowski, Logan, Kath, Riecke, Demb, and Singer]{jarsky_synaptic_2011}
Jarsky, T., Cembrowski, M., Logan, S.~M., Kath, W.~L., Riecke, H., Demb, J.~B., and Singer, J.~H.
\newblock A synaptic mechanism for retinal adaptation to luminance and contrast.
\newblock \emph{The Journal of Neuroscience: The Official Journal of the Society for Neuroscience}, 31\penalty0 (30):\penalty0 11003--11015, July 2011.
\newblock ISSN 1529-2401.
\newblock \doi{10.1523/JNEUROSCI.2631-11.2011}.

\bibitem[Wan and Heidelberger(2011)]{wan_synaptic_2011}
Wan, Q.-F. and Heidelberger, R.
\newblock Synaptic release at mammalian bipolar cell terminals.
\newblock \emph{Visual Neuroscience}, 28\penalty0 (1):\penalty0 109--119, January 2011.
\newblock ISSN 0952-5238, 1469-8714.
\newblock \doi{10.1017/S0952523810000453}.
\newblock Num Pages: 11 Place: New York Publisher: Cambridge Univ Press Web of Science ID: WOS:000286951300011.

\bibitem[Hosoya et~al.(2005)Hosoya, Baccus, and Meister]{hosoya_dynamic_2005}
Hosoya, T., Baccus, S.~A., and Meister, M.
\newblock Dynamic predictive coding by the retina.
\newblock \emph{Nature}, 436:\penalty0 71--7, July 2005.

\bibitem[Ebert et~al.(2024)Ebert, Buffet, Sermet, Marre, and Cessac]{ebert_temporal_2024}
Ebert, S., Buffet, T., Sermet, B.~S., Marre, O., and Cessac, B.
\newblock Temporal pattern recognition in retinal ganglion cells is mediated by dynamical inhibitory synapses.
\newblock \emph{Nature Communications}, 15\penalty0 (1):\penalty0 6118, July 2024.
\newblock ISSN 2041-1723.
\newblock \doi{10.1038/s41467-024-50506-7}.
\newblock URL \url{https://www.nature.com/articles/s41467-024-50506-7}.
\newblock Publisher: Nature Publishing Group.

\bibitem[Cybenko(1989)]{cybenko_approximation_1989}
Cybenko, G.
\newblock Approximation by superpositions of a sigmoidal function.
\newblock \emph{Mathematics of Control, Signals and Systems}, 2:\penalty0 303--314, 1989.
\newblock ISSN 0932-4194.

\bibitem[Hornik et~al.(1989)Hornik, Stinchcombe, and White]{hornik_multilayer_1989}
Hornik, K., Stinchcombe, M., and White, H.
\newblock Multilayer feedforward networks are universal approximators.
\newblock \emph{Neural Networks}, 2:\penalty0 359--366, 1989.

\bibitem[Euler et~al.(2002)Euler, Detwiler, and Denk]{eulerDirectionallySelectiveCalcium2002}
Euler, T., Detwiler, P.~B., and Denk, W.
\newblock Directionally selective calcium signals in dendrites of starburst amacrine cells.
\newblock \emph{Nature}, 418\penalty0 (6900):\penalty0 845--852, August 2002.
\newblock ISSN 0028-0836.
\newblock \doi{10.1038/nature00931}.

\bibitem[Asari and Meister(2012)]{asari_divergence_2012}
Asari, H. and Meister, M.
\newblock Divergence of visual channels in the inner retina.
\newblock \emph{Nat Neurosci}, 15:\penalty0 1581--9, October 2012.
\newblock \doi{10.1038/nn.3241}.

\bibitem[Matsumoto et~al.(2021)Matsumoto, Agbariah, Nolte, Andrawos, Levi, Sabbah, and Yonehara]{matsumoto_direction_2021}
Matsumoto, A., Agbariah, W., Nolte, S.~S., Andrawos, R., Levi, H., Sabbah, S., and Yonehara, K.
\newblock Direction selectivity in retinal bipolar cell axon terminals.
\newblock \emph{Neuron}, 109\penalty0 (18):\penalty0 2928--2942.e8, September 2021.
\newblock ISSN 0896-6273.
\newblock \doi{10.1016/j.neuron.2021.07.008}.
\newblock URL \url{https://www.sciencedirect.com/science/article/pii/S0896627321005183}.

\bibitem[Keat et~al.(2001)Keat, Reinagel, Reid, and Meister]{keat_predicting_2001}
Keat, J., Reinagel, P., Reid, R.~C., and Meister, M.
\newblock Predicting every spike: a model for the responses of visual neurons.
\newblock \emph{Neuron}, 30:\penalty0 803--17., 2001.

\bibitem[Pillow et~al.(2008)Pillow, Shlens, Paninski, Sher, Litke, Chichilnisky, and Simoncelli]{pillow_spatio-temporal_2008}
Pillow, J.~W., Shlens, J., Paninski, L., Sher, A., Litke, A.~M., Chichilnisky, E.~J., and Simoncelli, E.~P.
\newblock Spatio-temporal correlations and visual signalling in a complete neuronal population.
\newblock \emph{Nature}, 454:\penalty0 995--9, August 2008.
\newblock \doi{10.1038/nature07140}.

\bibitem[Qiu et~al.(2021)Qiu, Zhao, Klindt, Kautzky, Szatko, Schaeffel, Rifai, Franke, Busse, and Euler]{qiu_natural_2021}
Qiu, Y., Zhao, Z., Klindt, D., Kautzky, M., Szatko, K.~P., Schaeffel, F., Rifai, K., Franke, K., Busse, L., and Euler, T.
\newblock Natural environment statistics in the upper and lower visual field are reflected in mouse retinal specializations.
\newblock \emph{Current Biology}, 31\penalty0 (15):\penalty0 3233--+, August 2021.
\newblock ISSN 0960-9822, 1879-0445.
\newblock \doi{10.1016/j.cub.2021.05.017}.
\newblock Num Pages: 22 Place: Cambridge Publisher: Cell Press Web of Science ID: WOS:000685572700003.

\bibitem[Brackbill et~al.(2020)Brackbill, Rhoades, Kling, Shah, Sher, Litke, and Chichilnisky]{brackbill_reconstruction_2020}
Brackbill, N., Rhoades, C., Kling, A., Shah, N.~P., Sher, A., Litke, A.~M., and Chichilnisky, E.
\newblock Reconstruction of natural images from responses of primate retinal ganglion cells.
\newblock \emph{eLife}, 9:\penalty0 e58516, November 2020.
\newblock ISSN 2050-084X.
\newblock \doi{10.7554/eLife.58516}.
\newblock URL \url{https://doi.org/10.7554/eLife.58516}.
\newblock Publisher: eLife Sciences Publications, Ltd.

\bibitem[Karamanlis et~al.(2025)Karamanlis, Khani, Schreyer, Zapp, Mietsch, and Gollisch]{karamanlis_nonlinear_2025}
Karamanlis, D., Khani, M.~H., Schreyer, H.~M., Zapp, S.~J., Mietsch, M., and Gollisch, T.
\newblock Nonlinear receptive fields evoke redundant retinal coding of natural scenes.
\newblock \emph{Nature}, 637\penalty0 (8045):\penalty0 394--401, January 2025.
\newblock ISSN 1476-4687.
\newblock \doi{10.1038/s41586-024-08212-3}.
\newblock URL \url{https://www.nature.com/articles/s41586-024-08212-3}.
\newblock Publisher: Nature Publishing Group.

\bibitem[Maheswaranathan et~al.(2023)Maheswaranathan, McIntosh, Tanaka, Grant, Kastner, Melander, Nayebi, Brezovec, Wang, Ganguli, and Baccus]{maheswaranathan_interpreting_2023}
Maheswaranathan, N., McIntosh, L.~T., Tanaka, H., Grant, S., Kastner, D.~B., Melander, J.~B., Nayebi, A., Brezovec, L.~E., Wang, J.~H., Ganguli, S., and Baccus, S.~A.
\newblock Interpreting the retinal neural code for natural scenes: {From} computations to neurons.
\newblock \emph{Neuron}, 111\penalty0 (17):\penalty0 2742--2755.e4, September 2023.
\newblock ISSN 1097-4199.
\newblock \doi{10.1016/j.neuron.2023.06.007}.

\bibitem[Sanes and Masland(2015)]{sanes_types_2015}
Sanes, J.~R. and Masland, R.~H.
\newblock The types of retinal ganglion cells: current status and implications for neuronal classification.
\newblock \emph{Annu Rev Neurosci}, 38:\penalty0 221--46, July 2015.
\newblock \doi{10.1146/annurev-neuro-071714-034120}.

\bibitem[Gollisch and Meister(2010)]{gollisch_eye_2010}
Gollisch, T. and Meister, M.
\newblock Eye smarter than scientists believed: neural computations in circuits of the retina.
\newblock \emph{Neuron}, 65:\penalty0 150--64, January 2010.
\newblock \doi{10.1016/j.neuron.2009.12.009}.

\bibitem[Gollisch(2013)]{gollisch_features_2013}
Gollisch, T.
\newblock Features and functions of nonlinear spatial integration by retinal ganglion cells.
\newblock \emph{Journal of Physiology, Paris}, 107\penalty0 (5):\penalty0 338--348, November 2013.
\newblock ISSN 1769-7115.
\newblock \doi{10.1016/j.jphysparis.2012.12.001}.
\newblock URL \url{https://www.sciencedirect.com/science/article/pii/S0928425712000691}.

\bibitem[Kastner and Baccus(2014)]{kastner_insights_2014}
Kastner, D.~B. and Baccus, S.~A.
\newblock Insights from the retina into the diverse and general computations of adaptation, detection, and prediction.
\newblock \emph{Current Opinion in Neurobiology}, 25:\penalty0 63--69, April 2014.
\newblock ISSN 0959-4388.
\newblock \doi{10.1016/j.conb.2013.11.012}.
\newblock URL \url{https://www.sciencedirect.com/science/article/pii/S0959438813002237}.

\bibitem[Kerschensteiner(2022)]{kerschensteiner_feature_2022}
Kerschensteiner, D.
\newblock Feature {Detection} by {Retinal} {Ganglion} {Cells}.
\newblock \emph{Annual Review of Vision Science}, 8:\penalty0 135--169, 2022.
\newblock ISSN 2374-4642, 2374-4650.
\newblock \doi{10.1146/annurev-vision-100419-112009}.
\newblock Num Pages: 35 Place: Palo Alto Publisher: Annual Reviews Web of Science ID: WOS:000856252600006.

\bibitem[Turner et~al.(2018)Turner, Schwartz, and Rieke]{turner_receptive_2018}
Turner, M.~H., Schwartz, G.~W., and Rieke, F.
\newblock Receptive field center-surround interactions mediate context-dependent spatial contrast encoding in the retina.
\newblock \emph{eLife}, 7:\penalty0 e38841, September 2018.
\newblock ISSN 2050-084X.
\newblock \doi{10.7554/eLife.38841}.
\newblock URL \url{https://doi.org/10.7554/eLife.38841}.
\newblock Publisher: eLife Sciences Publications, Ltd.

\bibitem[Enroth-Cugell and Robson(1966)]{enroth-cugell_contrast_1966}
Enroth-Cugell, C. and Robson, J.~G.
\newblock The contrast sensitivity of retinal ganglion cells of the cat.
\newblock \emph{J Physiol}, 187:\penalty0 517--52, December 1966.

\bibitem[Krieger et~al.(2017)Krieger, Qiao, Rousso, Sanes, and Meister]{krieger_four_2017}
Krieger, B., Qiao, M., Rousso, D.~L., Sanes, J.~R., and Meister, M.
\newblock Four alpha ganglion cell types in mouse retina: {Function}, structure, and molecular signatures.
\newblock \emph{PLoS One}, 12:\penalty0 e0180091, 2017.
\newblock ISSN 1932-6203.
\newblock \doi{10.1371/journal.pone.0180091}.

\bibitem[Schwartz et~al.(2012)Schwartz, Okawa, Dunn, Morgan, Kerschensteiner, Wong, and Rieke]{schwartz_spatial_2012}
Schwartz, G.~W., Okawa, H., Dunn, F.~A., Morgan, J.~L., Kerschensteiner, D., Wong, R.~O., and Rieke, F.
\newblock The spatial structure of a nonlinear receptive field.
\newblock \emph{Nat Neurosci}, 15:\penalty0 1572--80, November 2012.
\newblock \doi{10.1038/nn.3225}.

\bibitem[Barlow and Levick(1965)]{barlowMechanismDirectionallySelective1965}
Barlow, H.~B. and Levick, W.~R.
\newblock The mechanism of directionally selective units in rabbit's retina.
\newblock \emph{J Physiol}, 178:\penalty0 477--504, June 1965.

\bibitem[Ölveczky et~al.(2003)Ölveczky, Baccus, and Meister]{olveczky_segregation_2003}
Ölveczky, B.~P., Baccus, S.~A., and Meister, M.
\newblock Segregation of object and background motion in the retina.
\newblock \emph{Nature}, 423:\penalty0 401--8, May 2003.
\newblock \doi{10.1038/nature01652}.

\bibitem[Jacoby and Schwartz(2017)]{jacoby_three_2017}
Jacoby, J. and Schwartz, G.~W.
\newblock Three {Small}-{Receptive}-{Field} {Ganglion} {Cells} in the {Mouse} {Retina} {Are} {Distinctly} {Tuned} to {Size}, {Speed}, and {Object} {Motion}.
\newblock \emph{Journal of Neuroscience}, 37\penalty0 (3):\penalty0 610--625, January 2017.
\newblock ISSN 0270-6474.
\newblock \doi{10.1523/JNEUROSCI.2804-16.2017}.
\newblock WOS:000393563900016.

\bibitem[Münch et~al.(2009)Münch, da~Silveira, Siegert, Viney, Awatramani, and Roska]{munch_approach_2009}
Münch, T.~A., da~Silveira, R.~A., Siegert, S., Viney, T.~J., Awatramani, G.~B., and Roska, B.
\newblock Approach sensitivity in the retina processed by a multifunctional neural circuit.
\newblock \emph{Nature Neuroscience}, 12\penalty0 (10):\penalty0 1308--1316, October 2009.
\newblock ISSN 1546-1726.
\newblock \doi{10.1038/nn.2389}.
\newblock URL \url{https://www.nature.com/articles/nn.2389}.
\newblock Publisher: Nature Publishing Group.

\bibitem[Appleby and Manookin(2020)]{appleby_selectivity_2020}
Appleby, T.~R. and Manookin, M.~B.
\newblock Selectivity to approaching motion in retinal inputs to the dorsal visual pathway.
\newblock \emph{eLife}, 9:\penalty0 e51144, February 2020.
\newblock ISSN 2050-084X.
\newblock \doi{10.7554/eLife.51144}.
\newblock URL \url{https://doi.org/10.7554/eLife.51144}.
\newblock Publisher: eLife Sciences Publications, Ltd.

\bibitem[Berry et~al.(1999)Berry, Brivanlou, Jordan, and Meister]{berry_anticipation_1999}
Berry, M.~J., Brivanlou, I.~H., Jordan, T.~A., and Meister, M.
\newblock Anticipation of moving stimuli by the retina.
\newblock \emph{Nature}, 398:\penalty0 334--8, March 1999.

\bibitem[Leonardo and Meister(2013)]{leonardo_nonlinear_2013}
Leonardo, A. and Meister, M.
\newblock Nonlinear dynamics support a linear population code in a retinal target-tracking circuit.
\newblock \emph{J Neurosci}, 33:\penalty0 16971--82, October 2013.
\newblock \doi{10.1523/JNEUROSCI.2257-13.2013}.

\bibitem[Johnston and Lagnado(2015)]{johnston_general_2015}
Johnston, J. and Lagnado, L.
\newblock General features of the retinal connectome determine the computation of motion anticipation.
\newblock \emph{Elife}, 4, 2015.
\newblock \doi{10.7554/eLife.06250}.

\bibitem[Gollisch and Meister(2008{\natexlab{b}})]{gollisch_rapid_2008}
Gollisch, T. and Meister, M.
\newblock Rapid neural coding in the retina with relative spike latencies.
\newblock \emph{Science}, 319:\penalty0 1108--11, February 2008{\natexlab{b}}.
\newblock \doi{10.1126/science.1149639}.

\bibitem[Chong et~al.(2020)Chong, Moroni, Wilson, Shoham, Panzeri, and Rinberg]{chong_manipulating_2020}
Chong, E., Moroni, M., Wilson, C., Shoham, S., Panzeri, S., and Rinberg, D.
\newblock Manipulating synthetic optogenetic odors reveals the coding logic of olfactory perception.
\newblock \emph{Science (New York, N.Y.)}, 368\penalty0 (6497):\penalty0 eaba2357, June 2020.
\newblock ISSN 0036-8075.
\newblock \doi{10.1126/science.aba2357}.
\newblock URL \url{https://pmc.ncbi.nlm.nih.gov/articles/PMC8237706/}.

\bibitem[Engel et~al.(1991)Engel, König, Kreiter, and Singer]{engel_interhemispheric_1991}
Engel, A.~K., König, P., Kreiter, A.~K., and Singer, W.
\newblock Interhemispheric {Synchronization} of {Oscillatory} {Neuronal} {Responses} in {Cat} {Visual} {Cortex}.
\newblock \emph{Science}, 252\penalty0 (5009):\penalty0 1177--1179, May 1991.
\newblock \doi{10.1126/science.252.5009.1177}.
\newblock URL \url{https://www.science.org/doi/10.1126/science.252.5009.1177}.
\newblock Publisher: American Association for the Advancement of Science.

\bibitem[Neuenschwander and Singer(1996)]{neuenschwander_long-range_1996}
Neuenschwander, S. and Singer, W.
\newblock Long-range synchronization of oscillatory light responses in the cat retina and lateral geniculate nucleus.
\newblock \emph{Nature}, 379\penalty0 (6567):\penalty0 728--732, February 1996.
\newblock ISSN 0028-0836.
\newblock \doi{10.1038/379728a0}.

\bibitem[Neuenschwander et~al.(2023)Neuenschwander, Rosso, Branco, Freitag, Tehovnik, Schmidt, and Baron]{neuenschwander_functional_2023}
Neuenschwander, S., Rosso, G., Branco, N., Freitag, F., Tehovnik, E.~J., Schmidt, K.~E., and Baron, J.
\newblock On the {Functional} {Role} of {Gamma} {Synchronization} in the {Retinogeniculate} {System} of the {Cat}.
\newblock \emph{Journal of Neuroscience}, 43\penalty0 (28):\penalty0 5204--5220, July 2023.
\newblock ISSN 0270-6474, 1529-2401.
\newblock \doi{10.1523/JNEUROSCI.1550-22.2023}.
\newblock URL \url{https://www.jneurosci.org/content/43/28/5204}.
\newblock Publisher: Society for Neuroscience Section: Research Articles.

\bibitem[Martin(1998)]{martin_colour_1998}
Martin, P.~R.
\newblock Colour processing in the primate retina: recent progress.
\newblock \emph{The Journal of Physiology}, 513\penalty0 (Pt 3):\penalty0 631--638, December 1998.
\newblock ISSN 0022-3751.
\newblock \doi{10.1111/j.1469-7793.1998.631ba.x}.
\newblock URL \url{https://pmc.ncbi.nlm.nih.gov/articles/PMC2231327/}.

\bibitem[Joesch and Meister(2016)]{joesch_neuronal_2016}
Joesch, M. and Meister, M.
\newblock A neuronal circuit for colour vision based on rod-cone opponency.
\newblock \emph{Nature}, 532:\penalty0 236--9, April 2016.
\newblock \doi{10.1038/nature17158}.

\bibitem[Oesch et~al.(2011)Oesch, Wade~Kothmann, and Diamond]{oesch_illuminating_2011}
Oesch, N.~W., Wade~Kothmann, W., and Diamond, J.~S.
\newblock Illuminating synapses and circuitry in the retina.
\newblock \emph{Current Opinion in Neurobiology}, 21\penalty0 (2):\penalty0 238--244, April 2011.
\newblock ISSN 0959-4388.
\newblock \doi{10.1016/j.conb.2011.01.008}.
\newblock URL \url{https://www.sciencedirect.com/science/article/pii/S0959438811000237}.

\bibitem[London and Hausser(2005)]{london_dendritic_2005}
London, M. and Hausser, M.
\newblock Dendritic computation.
\newblock \emph{Annu Rev Neurosci}, 28:\penalty0 503--32, 2005.
\newblock ISSN 0147-006X.
\newblock \doi{10.1146/annurev.neuro.28.061604.135703}.

\bibitem[Hartveit et~al.(2022)Hartveit, Veruki, and Zandt]{hartveit_dendritic_2022}
Hartveit, E., Veruki, M.~L., and Zandt, B.-J.
\newblock Dendritic {Morphology} of an {Inhibitory} {Retinal} {Interneuron} {Enables} {Simultaneous} {Local} and {Global} {Synaptic} {Integration}.
\newblock \emph{Journal of Neuroscience}, 42\penalty0 (9):\penalty0 1630--1647, March 2022.
\newblock ISSN 0270-6474, 1529-2401.
\newblock \doi{10.1523/JNEUROSCI.0695-21.2021}.
\newblock URL \url{https://www.jneurosci.org/content/42/9/1630}.
\newblock Publisher: Society for Neuroscience Section: Research Articles.

\bibitem[Cunningham and Yu(2014)]{cunningham_dimensionality_2014}
Cunningham, J.~P. and Yu, B.~M.
\newblock Dimensionality reduction for large-scale neural recordings.
\newblock \emph{Nature Neuroscience}, 17\penalty0 (11):\penalty0 1500--1509, November 2014.
\newblock ISSN 1546-1726.
\newblock \doi{10.1038/nn.3776}.

\bibitem[Barlow(1961)]{barlowPossiblePrinciplesUnderlying1961}
Barlow, H.~B.
\newblock Possible principles underlying the transformations of sensory messages.
\newblock In Rosenblith, W.~A., editor, \emph{Sensory {{Communication}}}, pages 217--234. MIT Press, Cambridge, MA, 1961.

\bibitem[Simoncelli(2003)]{simoncelli_vision_2003}
Simoncelli, E.~P.
\newblock Vision and the statistics of the visual environment.
\newblock \emph{Current Opinion in Neurobiology}, 13\penalty0 (2):\penalty0 144--149, April 2003.
\newblock ISSN 0959-4388.
\newblock \doi{10.1016/S0959-4388(03)00047-3}.
\newblock URL \url{https://www.sciencedirect.com/science/article/pii/S0959438803000473}.

\bibitem[Manookin and Rieke(2023)]{manookin_two_2023}
Manookin, M.~B. and Rieke, F.
\newblock Two {Sides} of the {Same} {Coin}: {Efficient} and {Predictive} {Neural} {Coding}.
\newblock \emph{Annual Review of Vision Science}, 9:\penalty0 293--311, September 2023.
\newblock ISSN 2374-4650.
\newblock \doi{10.1146/annurev-vision-112122-020941}.

\bibitem[Zheng and Meister(2025)]{zheng_unbearable_2025}
Zheng, J. and Meister, M.
\newblock The unbearable slowness of being: {Why} do we live at 10 bits/s?
\newblock \emph{Neuron}, 113\penalty0 (2):\penalty0 192--204, January 2025.
\newblock ISSN 0896-6273.
\newblock \doi{10.1016/j.neuron.2024.11.008}.
\newblock URL \url{https://www.cell.com/neuron/abstract/S0896-6273(24)00808-0}.
\newblock Publisher: Elsevier.

\bibitem[Yonehara et~al.(2009)Yonehara, Ishikane, Sakuta, Shintani, Nakamura-Yonehara, Kamiji, Usui, and Noda]{yonehara_identification_2009}
Yonehara, K., Ishikane, H., Sakuta, H., Shintani, T., Nakamura-Yonehara, K., Kamiji, N.~L., Usui, S., and Noda, M.
\newblock Identification of retinal ganglion cells and their projections involved in central transmission of information about upward and downward image motion.
\newblock \emph{PLoS ONE}, 4\penalty0 (1):\penalty0 e4320, 2009.
\newblock ISSN 1932-6203.
\newblock \doi{10.1371/journal.pone.0004320}.
\newblock URL \url{http://journals.plos.org/plosone/article?id=10.1371/journal.pone.0004320}.

\bibitem[Yilmaz and Meister(2013)]{yilmaz_rapid_2013}
Yilmaz, M. and Meister, M.
\newblock Rapid innate defensive responses of mice to looming visual stimuli.
\newblock \emph{Curr Biol}, 23:\penalty0 2011--5, October 2013.
\newblock ISSN 1879-0445 (Electronic) 0960-9822 (Linking).
\newblock \doi{10.1016/j.cub.2013.08.015}.

\bibitem[Kim et~al.(2020)Kim, Shen, Hsiang, Johnson, and Kerschensteiner]{kim_dendritic_2020}
Kim, T., Shen, N., Hsiang, J.-C., Johnson, K., and Kerschensteiner, D.
\newblock Dendritic and parallel processing of visual threats in the retina control defensive responses.
\newblock \emph{Science Advances}, 6\penalty0 (47):\penalty0 eabc9920, November 2020.
\newblock \doi{10.1126/sciadv.abc9920}.
\newblock URL \url{https://www.science.org/doi/10.1126/sciadv.abc9920}.
\newblock Publisher: American Association for the Advancement of Science.

\bibitem[Wang et~al.(2021)Wang, Li, De, Wu, and Zhang]{wang_off-transient_2021}
Wang, F., Li, E., De, L., Wu, Q., and Zhang, Y.
\newblock {OFF}-transient alpha {RGCs} mediate looming triggered innate defensive response.
\newblock \emph{Current Biology}, 0\penalty0 (0), April 2021.
\newblock ISSN 0960-9822.
\newblock \doi{10.1016/j.cub.2021.03.025}.
\newblock URL \url{https://www.cell.com/current-biology/abstract/S0960-9822(21)00364-X}.
\newblock Publisher: Elsevier.

\bibitem[Reinhard et~al.(2019)Reinhard, Li, Do, Burke, Heynderickx, and Farrow]{reinhard_projection_2019}
Reinhard, K., Li, C., Do, Q., Burke, E.~G., Heynderickx, S., and Farrow, K.
\newblock A projection specific logic to sampling visual inputs in mouse superior colliculus.
\newblock \emph{eLife}, 8:\penalty0 e50697, November 2019.
\newblock ISSN 2050-084X.
\newblock \doi{10.7554/eLife.50697}.
\newblock URL \url{https://doi.org/10.7554/eLife.50697}.
\newblock Publisher: eLife Sciences Publications, Ltd.

\bibitem[Hoy et~al.(2019)Hoy, Bishop, and Niell]{hoy_defined_2019}
Hoy, J.~L., Bishop, H.~I., and Niell, C.~M.
\newblock Defined {Cell} {Types} in {Superior} {Colliculus} {Make} {Distinct} {Contributions} to {Prey} {Capture} {Behavior} in the {Mouse}.
\newblock \emph{Current biology: CB}, 29\penalty0 (23):\penalty0 4130--4138.e5, December 2019.
\newblock ISSN 1879-0445.
\newblock \doi{10.1016/j.cub.2019.10.017}.
\newblock URL \url{https://www.cell.com/current-biology/abstract/S0960-9822(19)31323-5}.

\end{thebibliography}

\end{document}